# Evidence of orbital reconstruction at interfaces in $La_{0.67}Sr_{0.33}MnO_3$ films

A Tebano, C. Aruta, S. Sanna, P.G. Medaglia, and G. Balestrino

*CNR-INFM Coherentia and Dipartimento di Ingegneria Meccanica, Università di Roma Tor Vergata, Via del Politecnico 1, 00133 Roma, Italy*

A.A. Sidorenko, R. De Renzi

*Dipartimento di Fisica and Unità CNISM di Parma, Via G.P. Usberti 7A, I-43100 Parma, Italy*

G. Ghiringhelli, L. Braicovich

*CNR-INFM Coherentia and Soft, Dipartimento di Fisica, Politecnico di Milano, piazza Leonardo da Vinci 32, I-20133 Milano, Italy*

V. Bisogni, N.B. Brookes

*European Synchrotron Radiation Facility, Boîte Postale 220, 38043 Grenoble, France*

## ABSTRACT

Linear dichroism (LD) in x-ray absorption, diffraction, transport and magnetization measurements on thin $La_{0.7}Sr_{0.3}MnO_3$ films grown on different substrates, allow identification of a peculiar interface effect, related just to the presence of the interface. We report the LD signature of preferential *$3d$-$e_g$($3z^2$-$r^2$)* occupation at the interface, suppressing the double exchange mechanism.

This surface orbital reconstruction is opposite to that favored by residual strain and is independent of dipolar fields, the chemical nature of the substrate and the presence of capping layers.

Interfaces between perovskite oxides display unexpected properties. Recent notable examples include insulating, non magnetic multilayers, like $SrTiO_3$ (STO) and $LaAlO_3$ (LAO) [1], giving rise to high conductivity [2] and possibly magnetism and superconductivity [3]. Oxygen vacancies [4], epitaxial strain, polarity discontinuities from interface-generated dipoles [5] and orbital reconstruction at the interface [6, 7] play a role still not understood. An important case is the interface between manganite and STO, providing tunnel junctions for manganite-based devices, such as spin valves or spin injectors with a high degree of spin polarization and robust magnetic properties at the interface between the manganite and the barrier. Several authors found that the double exchange (DE) interface magneto-transport properties are strongly depressed below a critical thickness value $t_c$ which depends on the specific substrate and differs with experimental conditions [8,9,10] ranging from 3 nm for $La_{0.7}Sr_{0.3}MnO_3$ (LSMO) films grown on (110) oriented $NdGaO_3$ (NGO) substrates to 15 nm for LAO substrates.

The roles of chemistry, polarization and strain may be singled out by selective experiments, e.g. Ref. [11], where an engineered interface obtained by intercalating two LMO unit cells (u.c.) between the LSMO and the STO has been shown to recover the LSMO bulk properties even at room temperature. The role of strain on preferential orbital occupation in transition metal oxides has been widely studied[12]. The anisotropy of $d$-orbitals influences the electron correlation effects in an orbital direction-dependent manner, thus giving rise to the anisotropy of the electron-transfer and eventually destroying the DE order of unstrained half-metallic LSMO (Fig.1, center). The strain effect on orbital physics can be understood on the basis of the experimental phase diagram proposed by Konishi *et al.*, [13] and explained theoretically by Fang *et al.*[14]. Spin ordering in strained manganite is influenced by orbital ordering and several anti-ferromagnetic (AF) insulating Jahn-Teller distorted phases are observed: the strain induced elongation or compression of the $MnO_6$ octahedra leads to crystal field splitting of the $e_g$ levels, lowering either *i)* the $(3z^2-r^2)$ state which favors the C-type AF structure (Fig.1, left) or *ii)* the $(x^2-y^2)$ state which stabilizes the A-type structure (Fig.1, right). The linear dichroism of the x-ray absorption (LD-XAS) at the Mn-$L_{2,3}$ edge is a powerful and direct probe of the $3d$ orbital occupation in manganites [15]. LD-XAS gave recently strong support for case *i)* [16], showing that the signature of $(3z^2-r^2)$ orbital occupancy persists in partially unrelaxed and compressed in plane LSMO films grown epitaxially on LAO. A different mechanism must originate the suppression of interface DE magneto-transport properties in LSMO films grown on STO and on nearly lattice-matched NGO substrates. In this case both substrates induce weak tensile strain, which is insufficient to determine the transition from a DE ferromagnet to an A-type antiferromagnet [13,14] Chemical composition effects at the interface must be specifically addressed by controlling the deposition conditions, e.g. by trying to enhance them in ad-hoc experiments. Furthermore it has been theoretically considered that the presence of the surface/interface discontinuity affects the orbital occupation in ultra-thin manganite films, and thus their electronic properties, by the following mechanisms: *a)* reduced film thickness breaks the bulk degeneracy between the $3z^2$-$r^2$ and the $x^2$-$y^2$ bands, localizing the $3z^2$-$r^2$ electrons [6]; *b)* the shift of the surface electronic levels with respect to the bulk energy (it has been pointed out that, in the case of manganite films terminated by a *MnO* plane at the bare surface, the $e_g(3z^2$-$r^2)$ orbitals are energetically favored relative to the $e_g(x^2$-$y^2)$ orbitals [17]); *c)* electrical charge unbalance at the interface, induced by the substrate termination or by a suitable buffer layer[7,18].

In this letter we report on LSMO ultra-thin layers and heterostructures on different substrates, with and without capping layers, designed to correlate the depression of the DE phase with strain. Fully oxidized manganite films, having thickness ranging from 6 u.c. to 100 u.c., were grown by PLD with *in-situ* Reflection High Energy Electron Diffraction diagnostic, allowing us to calibrate film thickness with the precision of a single unit cell. STO substrates were etched in order to have an ideal $TiO_2$ surface termination [19]. Film structural properties were investigated by x-ray diffraction measurements to obtain the lattice parameters. Electrical transport measurements were carried out by the standard four-probe technique as a function of temperature.

XAS measurements were performed at the beam line ID08 of the European Synchrotron Radiation Facility (Grenoble). Using linearly polarized radiation, we measured the XAS at the Mn-$L_{2,3}$ edge, corresponding to the 2p ? 3d resonant transition. Polarization effects arise when the polarization vector is set parallel to the *c* crystallographic axis or perpendicular to it ($I_c$ and $I_{ab}$ respectively). The LD is the difference between the two spectra ($I_{ab}$-$I_c$) and gives a direct insight of the empty Mn *3d* states: a LD which is on average positive (negative) indicates a majority of off-plane (in-plane) empty *3d* states. Considering the crystal field splitting, the effect can be mainly related to the occupation of the two $e_g$ states ($3r^2$-$z^2$ and $x^2$-$y^2$) with majority spin: a LD which is on average positive (negative) is due to a preferential occupation of the in-plane $x^2$-$y^2$ (out-of-plane $3r^2$-$z^2$) orbital.

Magnetization measurements were carried out by a SQUID magnetometer. Further experimental details are given in ref. [16] and [26].

In Fig.2 the metal-insulator transition temperature ($T_P$) is plotted vs. film thickness for samples grown on three different substrates (STO, LAO and NGO), showing that suppression of magneto-transport properties occurs below 30 u.c. for films grown on LAO, where the effect originates from in-plane compressive strain [16], whereas $t_c$ is much shorter (about 7 u.c., within our experimental error of one unit cell) for films grown both on STO and NGO. In the case of STO, $t_c$ is by far shorter than the range of the unrelaxed weak tensile strain, in agreement with the strain being too weak to have an influence [11].

Additional information is provided by the temperature dependence of the magnetization, M(T), of the LSMO/STO films, shown in Fig.3: the Curie temperature ($T_C$) is strongly depressed with decreasing thickness but at least a partial ferromagnetic character is preserved even for t<$t_c$. Furthermore low temperature $^{55}$Mn NMR measurements [20, 21] have clearly shown that the NMR signal intensity from Mn DE vanishes with thickness around t=$t_c$ whereas a further ferromagnetic contribution from $Mn^{4+}$ survives, indicating the presence of a non-DE ferromagnetic phase with $Mn^{4+}$ / $Mn^{3+}$ charge separation. The presence of this ferromagnetic contribution is confirmed by the non-vanishing average saturation magnetization values reported in Fig.3 for thickness comparable with $t_c$ .

Fig.4(a) shows our key results: the LD-XAS displays a marked change, in LSMO/STO films, when crossing $t_c$. The spectra calculated for the two distinct $e_g$ orbital occupations, $3z^2-r^2$ (triangles) and $x^2-y^2$ (full circles) [22,23], are plotted in Fig.4(b), revealing opposite signs for these two cases. Although the comparison with experiments can only be qualitative and a proper fit is not feasible, the sign reversal is observed in the experimental spectra of Fig.4(a) for energies above the E 644 eV marker. In addition, the sign alternation in the peaks of the red and blue experimental curves qualitatively follows the predicted behavior of Fig.4(b), and that of analogous simulations, e.g. Fig.1 in ref. [15] and Fig.6 in ref. [23]. We conclude that well above $t_c$, (t=50 u.c.) LD-XAS shows the fingerprint of a preferential $(x^2-y^2)$ orbital ordering whereas below $t_c$ (t=6 u.c.), it shows a $(3z^2-r^2)$ preferential occupation also evident in the case of LSMO/NGO films (Fig.5), for t= $t_c$ .

This orbital occupancy change is even more striking when compared with the preferential $(x^2-y^2)$ character of relatively thick (100 u.c.) fully strained LSMO/STO films ($c/a$=0.98). These films display metallic ferromagnetic properties with $T_C$ (~ $T_P$)=370 K identical to the bulk value and their preferential orbital occupation is once more determined [22] from the LD-XAS spectra. This means that a small $(x^2-y^2)$-type LD is compatible with DE.

We argue that the sharp change shown in Fig.4(a) is related to the interface rather than to intensive film properties. However this feature could be attributed to two distinct mechanisms: symmetry breaking at the interface and at the free surface [6,7] or charge redistribution caused by localized oxygen deficiency [4]. Both these effects could lead to a decrease of the DE carriers in the manganite layer and thus to the deterioration of their magnetotransport properties.

We can rule out the chemical mechanism. STO and NGO have very different chemical properties (SrO and $TiO_2$ layers are neutral and could produce a polarization discontinuity [5] because of the LaO and $MnO_2$ layers charge unbalance, whereas NdO and $GaO_2$ layers are themselves charge unbalanced and do not provide this effect). In particular, PLD on STO is known to sputter oxygen from the substrate, which then acts as an oxygen getter on the first LSMO layers, but only when they are deposited in high vacuum conditions ($p<10^{-5}$ mbar, [24]). By contrast our films are deposited at higher oxygen pressure (average pressure $p=10^{-3}$ mbar, with 12% of ozone), which, with the low deposition rate, grants the correct oxygen stoichiometry [25]. Furthermore, recent LD XAS measurements carried out on oxygen deficient manganites have shown that DE carriers depletion favors the $(x^2-y^2)$ rather than the $(3z^2-r^2)$ orbital preferential occupation [26] hence any residual oxygen deficiency would produce a LD-XAS signature opposite to that observed. Finally, we have grown two series of heterostructures, LSMO/STO (a free surface plus a STO/LSMO interface) and STO/LSMO/STO (two STO/LSMO interfaces, with cap layer thickness of 10 u.c). $T_P$ values for these samples are shown in the inset of Fig.2 vs. the LSMO film thickness. The difference between the two data sets is below the error-bar. Also the magnetic properties ($T_C$ and saturation magnetization) of 4 nm LSMO films are not affected by a STO cap layer [27]. This implies that the free surface and the STO interface have the same effect, confirming that STO is playing no specific chemical role.

By exclusion we are left with only the symmetry-breaking effects as a possible explanation of our experimental results. The more likely physical mechanism linking the detected $(3z^2-r^2)$ preferential occupation at the interface/surface and the deterioration of the magneto-transport properties is that the occupation of $(3z^2-r^2)$ orbitals, even when induced by symmetry breaking and not by strain, tends to enhance the ferromagnetic coupling along the $c$ axis and the in-plane AF interaction between adjacent Mn cations [28]. Meanwhile depletion of the $(x^2-y^2)$ orbitals results in a weakening of the ferromagnetic DE interaction which is mediated by degenerate $e_g$ electrons. The cooperative effect of a ferromagnetic coupling along the $c$ axis and an in-plane AF coupling could result in a C-type AF ordering (Fig.1 left) in a thin layer at the surface/interface. Although we have no means of directly monitoring the magnetic structure prediction, the C-type phase is insulating and it agrees with our findings: on the basis of the transport measurements the thickness of such a surface/interface layer can be estimated as 3-4 u.c. (about half of $t_c$).

Since NMR shows that a residual ferromagnetic charge-localized phase is always present in films thinner than $t_c$ both on STO [22] and LAO [16], the picture must be slightly more complicated, leaving room for additional phase separated components when the plain DE mechanism is disrupted. This consideration suggests a rather blurred transition between the phases with insulating and half-metallic character, which agrees with the experimental observation (see also [16]) that the LD-XAS signal intensity (Fig.4(a)), corresponding to an effective exponentially weighted average over a decade of u.c., is smaller than that calculated (Fig.4(b)). The experimental spectra are probably weighted averages of spectra, with opposite LD and slightly different spectral parameters Dt and Ds. This justifies both the approximate match between experiment and simulation, as well as our use of the term *preferential* orbital occupation.

In conclusion, we have shown that, while in the case of LSMO grown on LAO, the worsening of the DE magnetotransport properties is a *bulk* effect, caused by strong in-plane compressive epitaxial strain, the same phenomenon in the case of the ultrathin LSMO films grown on STO (weak tensile strain) and NGO (almost unstrained) is of interfacial origin. Furthermore, transport and LD-XAS measurements have demonstrated that orbital reorganization is independent of the specific nature of the interface, either a free surface, LSMO/STO or LSMO/NGO interface. Such a result proves that the suppression of DE mechanism is not a consequence of a polarity discontinuity (which occurs for LSMO/STO but not for LSMO/NGO) but rather a consequence of the broken symmetry at the interfaces, which stabilizes the $3z^2-r^2$ against the $x^2-y^2$ orbitals.

**Acknowledgements**

We thank, Elettronica Masel s.r.l., for a partial financial support. Some of us (A.A.S. and R.D.R.) acknowledge support from STREP OFSPIN.

Figure captions

Fig. 1) (Color on line) Relation between strain and Mn orbital order: Left, in-plane compressive strain favors the $(3z^2-r^2)$ orbital and the C-type AF structure; Center, undistorted DE ferromagnetic structure; Right, tensile strain favors the $(x^2-y^2)$ orbital and the A-type AF structure.

Fig. 2) (Color on line) Behavior of the metal-insulator transition temperature ($T_P$) vs. thickness for films grown on different substrates, in the inset the behavior of $T_P$ vs. thickness for films grown on STO substrates is reported. Triangles refer to films capped by a STO layer (STO/LSMO/STO heterostructures), while full dots to uncapped films.

Fig. 3) (Color on line) Magnetization vs. temperature for four films grown on STO substrates having a thickness of 6 u.c. (triangles), 8 u.c. (squares), 12 u.c. (full dots) and 20 u.c. (diamonds). Saturation magnetization is obtained from independent M(H) curves.

Fig.4) (Color on line) (a):LD for films grown on STO substrates having a thickness of 6 u.c. (triangles), 10 u.c. (squares), 50 u.c. (full dots). LD is reported in percent of the XAS $L_3$ peak height. (b): Calculated LD in $Mn^{3+}$ $L_{2,3}$ absorption with 10Dq=1.1eV, Slater integrals reduced to 70%, T=300 K, interatomic exchange interaction=10meV, distortion octahedral parameters Dt=Ds/4 and different Ds: -0.186 eV (full dots) and 0.186 eV (triangles) [22,23]. (c): LD for two films grown on NGO substrates having a thickness of 6 u.c. (triangles) and 9 u.c. (squares). LD is reported in percent of the XAS $L_3$ peak height.

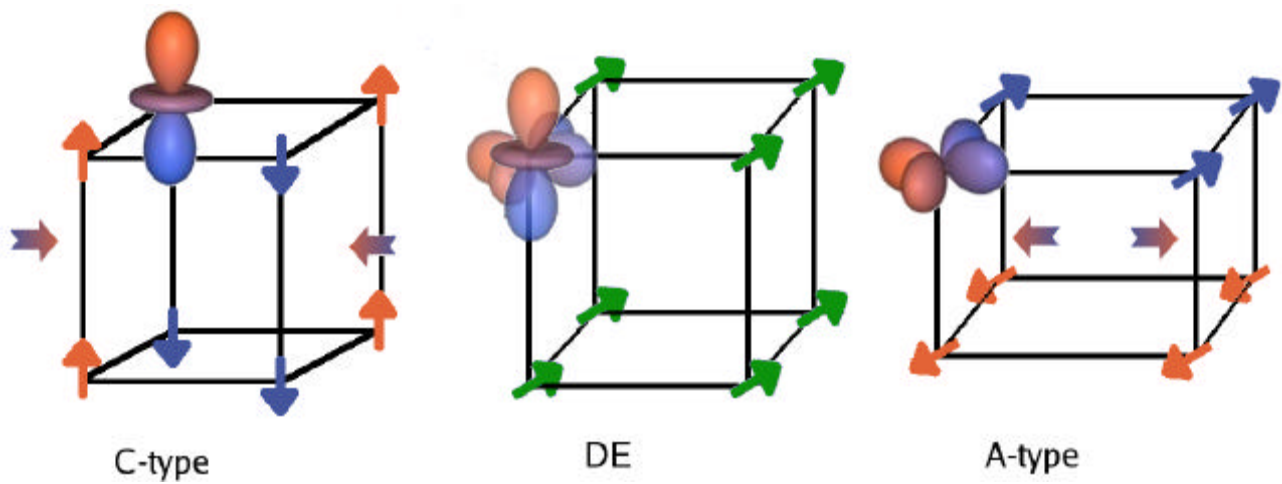

Fig1

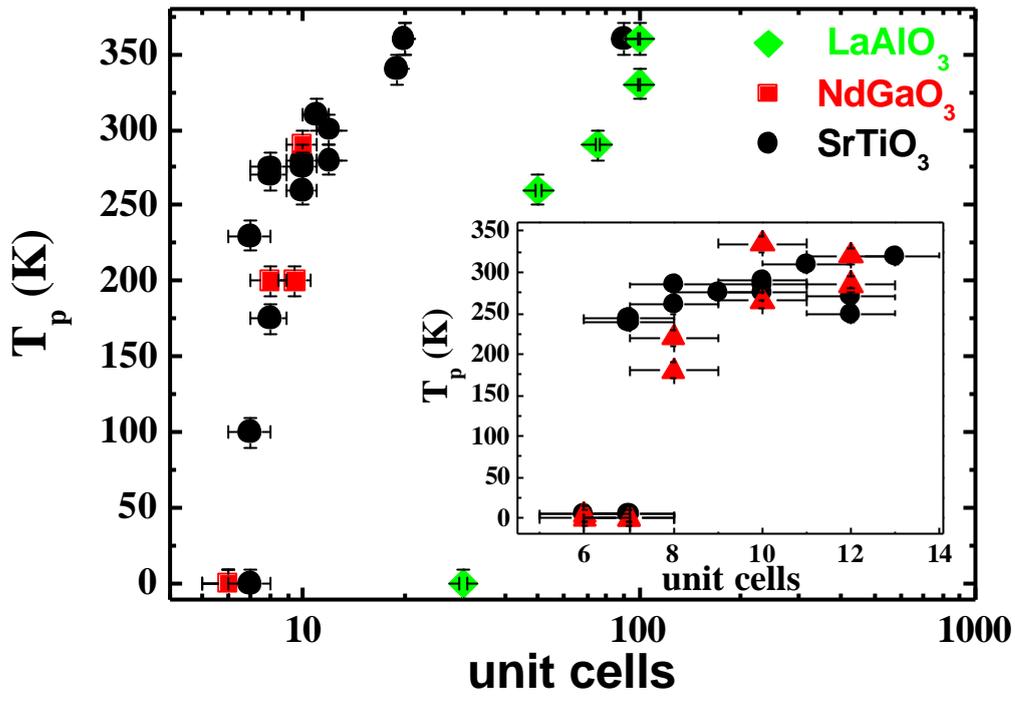

Fig. 2

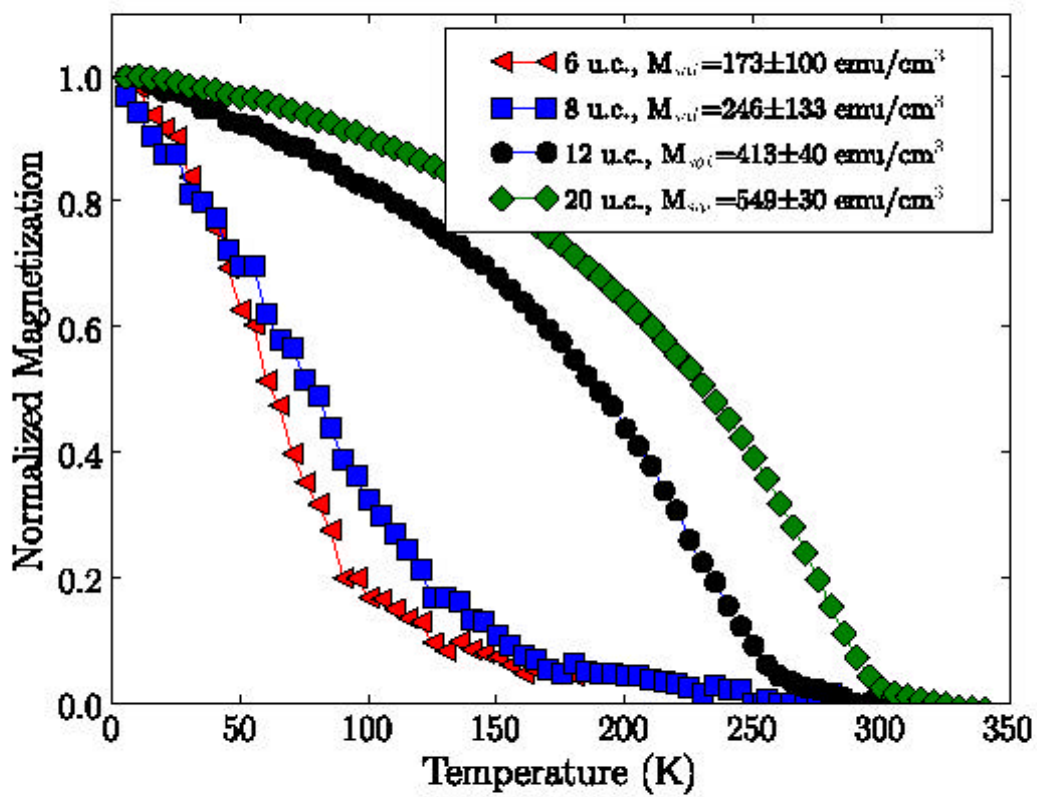

Fig. 3

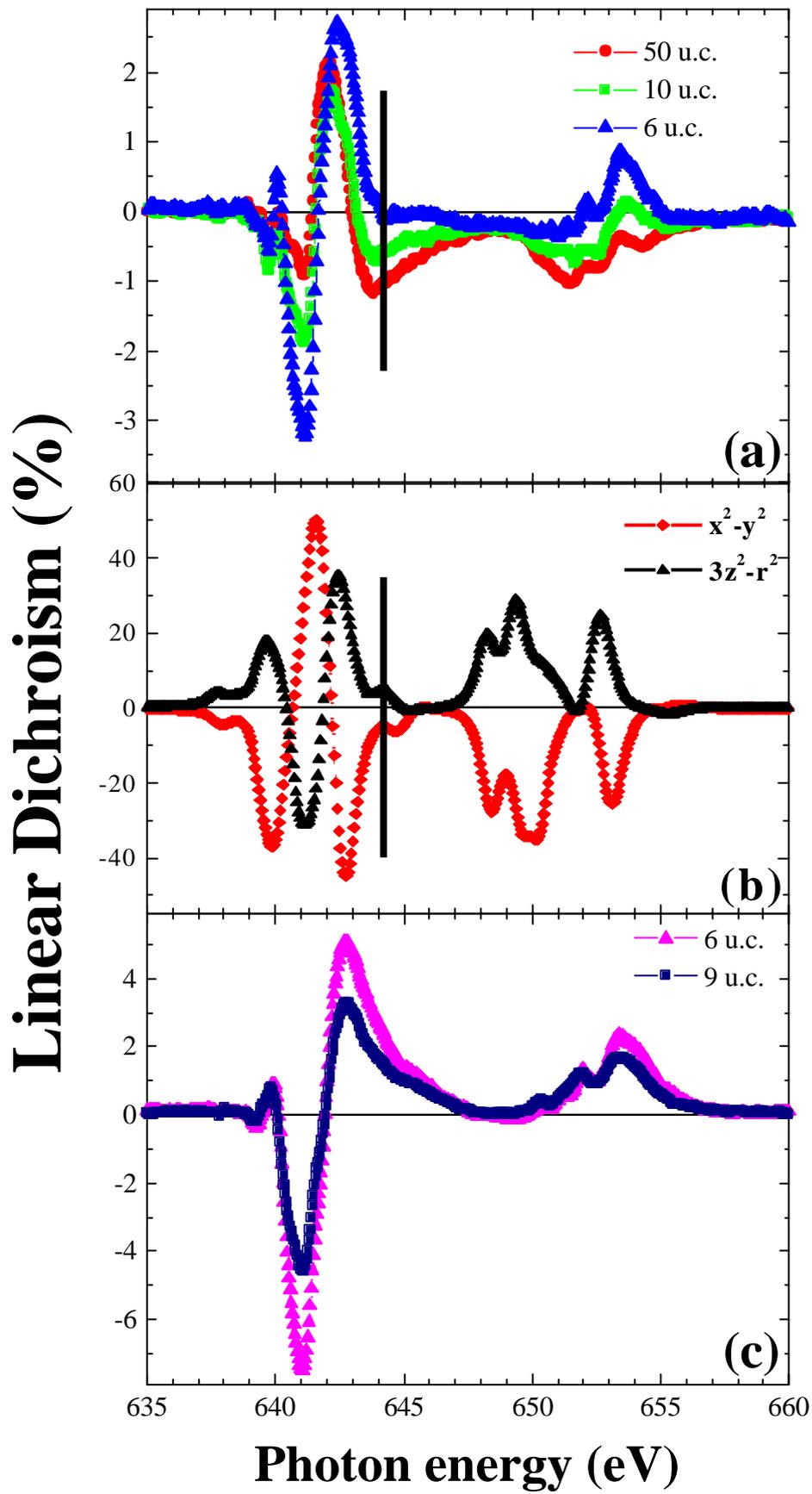

Fig. 4 (a,b and c)